\documentclass[12pt,preprint]{aastex}

\def\dem71{DEM L 71\,}

\newcommand{\be}{\begin{equation}}
\newcommand{\ee}{\end{equation}}
\newcommand{\nn}{\mbox{} \nonumber \\ \mbox{} }
\newcommand{\ba}{\begin{eqnarray}}
\newcommand{\ea}{\end{eqnarray}}

\newcommand{\Bf}{{magnetic field}}

\newcommand{\Ef}{{electric field}}

\newcommand{\E}{{\bf E}}
\newcommand{\B}{{\bf B}}

\newcommand{\J}{{\bf J}}
\renewcommand{\v}{{\bf v}}

\begin{document}

\title{In Situ Heating of the 2007 May 19 CME Ejecta Detected by STEREO/PLASTIC and ACE}
\author{Cara E. Rakowski,\altaffilmark1 J. Martin Laming\altaffilmark2 \&
Maxim Lyutikov\altaffilmark3}

\altaffiltext{1}{Space Science Division, Naval Research Laboratory, Code 7671, Washington DC 20375}
\altaffiltext{2}{Space Science Division, Naval Research Laboratory, Code 7674L, Washington DC 20375}
\altaffiltext{3}{Department of Physics, Purdue University, West Lafayette, IN 47907}

\begin{abstract}
{\it In situ} measurements of ion charge states can provide unique
insight into the heating and evolution of coronal mass ejections when
tested against realistic non-equilibrium ionization modeling. In
this work we investigate the representation of the CME magnetic
field as an expanding spheromak configuration, where the plasma heating is
prescribed by the choice of anomalous resistivity and the spheromak
dynamics. We chose as a test case, the 19 May 2007 CME observed by
STEREO and ACE.  The spheromak is an appealing physical model,
because the location and degree of heating is fixed by the choice of 
anomalous resistivity and the spheromak
expansion rate which we constrain with observations.
This model can provide the heating required
between 1.1$R_{\sun}$ and earth orbit to produce charge states
observed in the CME flux rope.


However this source of heating in the
spheromak alone has difficulty accounting for the rapid heating to
Fe$^{8 - 11+}$ at lower heights, as observed in STEREO EUVI due to the rapid radiative cooling that occurs at the high densities involved. Episodes of heating and cooling clearly unrelated to spheromak expansion are observed prior to the eruption, and presumably still play a role during the eruption itself. Spheromak heating is also not capable of reproducing the
high Fe charge states (Fe$^{16+}$ and higher) seen {\it in situ} exterior to the flux rope in this CME. Thus while the spheromak
configuration may be a valid model for the magnetic topology,
other means of energization are still required to provide much of the rapid heating observed.

\end{abstract}

\section{Introduction}

The eruption mechanism of coronal mass ejections (CMEs) is currently
an extremely active area of research. A variety of mechanisms have
been discussed in the literature. Most of these require some form of
magnetic reconnection during and after the eruption. The associated
thermal heating should be an important diagnostic of the eruptive
process, including the eruption dynamics and ejecta geometry, but
until now has been relatively unexploited.  Such heating gives rise to
UV-X-ray emissions during the eruption, that can be remotely sensed,
and also leaves its imprint in the charge states of ions detected {\it
  in situ} near 1 AU, as their ionization state responds to the newly
heated plasma. These techniques give complementary insights into the
heating and evolution. Spectroscopy is in some sense more direct,
being a prompt signature of the eruption, but with existing
instrumentation has only been possible relatively close to the solar
surface where the signal is strong. Ion charge states, on the other
hand, are routinely collected and CMEs, and their constituent parts
(forward shock, flux rope, etc) can be easily identified. Further
interpretation of such data require modeling of the non-equilibrium
ionization (NEI) balance as the CME plasma expands outward into the
ambient solar wind. In this paper we explore the consequences of a
different source of thermal heating, that due to anomalous resistivity
within a spheromak solution for the CME geometry as it expands into
the solar wind. We compare the predictions of such a model to STEREO
and ACE observations of the 19 May 2007 CME. The structure of the
paper is as follows. We
describe the status of our understanding of ion charge states observed
in CMEs in more detail in section 2. Section 3 describes the 2007 May
19 CME, and some of our motivation for exploring a spheromak
model. Section 4 gives the theory of spheromak resistive expansion,
with some discussion of just how closely the spheromak model used here
may represent a real CME. Section 5 outlines the NEI model
incorporating the spheromak resistive heating, and we conclude in section 6
with a summary of our results in the context of previous research and
discuss future modeling improvements.

\section{NEI modeling of CME charge states}

Interplanetary CMEs or ICMEs have many well-known signatures. Of
these, {\it in situ} measurements of ion charge state distributions
hold a unique potential for diagnosing the conditions throughout the
eruption. The charge states evolve through ionizations and
recombinations as a function of temperature and density up to heights
of 3 to 6 solar radii heliocentric distance but are ``frozen in''
thereafter. Since low-Z ions ``freeze-in'' at higher densities
(i.e. smaller heights or radii) than high-Z ions, they provide
complimentary constraints on the temperature evolution.

A number of studies have examined the existing solar wind composition
data and made inferences on freeze-in temperatures based on computed
ionization distributions appropriate to coronal equilibrium
(e.g. Zurbuchen and Richardson 2006, Zurbuchen 2004, Lepri and
Zurbuchen 2004, Lepri et al. 2001, Henke et al. 2001, Gloeckler et
al. 1998, Henke et al. 1998). However \citet{RLL07} were the first to
conduct time-dependent modeling of the ion charge state distributions
of various elements detected {\it in situ} in order to draw
quantitative conclusions regarding the thermal input and initial
conditions in the corona during the CME eruption. \citet{RLL07}
modeled the charge states of 8 ICMEs from the event list of
\citet{lynch03}, chosen to cover a range of velocities.

\citet{RLL07} considered a simplified
geometry for the CME ejecta and parameterized the evolution. The model velocity
and expansion were based on the observed phenomenology from plane-of-the-sky
events showing that typically CME height-time evolution consists of
three phases: initiation, acceleration and propagation
\citep{zhang01,zhang04,zhang06,sui03,lin05,sui05}.
Heating of the CME core plasma during the acceleration
phase was explored assuming a heating rate for the CME plasma
proportional to the rate of kinetic energy increase, i.e. a constant
fraction $QE/KE$ during the acceleration up to a final velocity $v_{f}$.
In 5 out of 8 sample CMEs studied, the dominant Fe charge states were
neon-like (16+) or higher, indicating that high temperatures,
comparable to flares ($\sim 10^7$K), are involved. Starting the plasma
from this temperature and allowing the ions to recombine as they
expand could often account for the Fe ionization balance, with peaks
around Fe$^{16+}$ and Fe$^{8+}$ (the Ne-like and Ar-like charge
states, which have small recombination rates to the next charge states
down, hence population ``bottlenecks'' here). However the lower-Z
elements placed a limit on the maximum starting temperature (at least
if assuming ionization equilibrium in the seed plasma). Above $\sim
2.5\times 10^{6}$K, O would have been mainly O$^{8+}$ instead of
O$^{7+}$ and O$^{6+}$ as observed and would not recombine
significantly during the CME evolution. Evidently plasma must start
out much cooler, and be further heated as the CME accelerates.
Thermal energy inputs on the order of 2 to 10 times the kinetic energy
were needed in 7 out of 8 events studied.

Similar conclusions to \citet{RLL07} about the thermal energy input to
CMEs have been reached from analysis of ultra-violet spectra taken by
SOHO/UVCS. \citet{akmal01} studied a 480 km s$^{-1}$ CME observed on
1999 April 23, and find a thermal energy comparable to the bulk
kinetic energy of the plasma. \citet{ciaravella01} give similar
results for the 260 km s$^{-1}$ 1997 December 12 CME. More
dramatically, \citet{lee09} studied the 2001 December 13 event, and
using the combination of [O V] density diagnostics and non-equilibrium
ionization modeling of O VI found that 75\% of the magnetic energy
must go into heat to match the UVCS observations. \citet{RLL07}
found a heat to kinetic energy ratio of at least 6 for the same
event.

The motivations behind the current work are twofold. First, to test if our
non-equilibrium ionization modeling of the charge state distributions can hold up
to the scrutiny of additional constraints on the eruption dynamics coming
from STEREO observations in the 2007 May 19 CME, as
well as observations regarding the charge states closer to the
sun. Secondly we explore whether a more physically and theoretically
specified heating model when incorporated into our non-equilibrium
modeling can reproduce the charge states of a real event with
reasonable physical parameters.

\section{The 2007 May 19 CME}
The 2007 May 19 CME studied in this paper is qualitatively different
from those CMEs discussed above.
As one of the first major CME eruptions in the STEREO era the 19 May
2007 event has been intensely studied by multiple authors
\citep{Gopalswamy, Veronig, Li08, Liu08,Kilpua, Liewer, Bone,Kerdraon}.
It began as a filament eruption from the active region AR10956 and was
detected as an ICME by STEREO B, ACE and possibly STEREO A on 2007 May 22.
Multiple heating and cooling episodes were seen in the two days prior
to eruption \citep{Liewer, Bone, Li08}.  The total
unsigned flux declined 17\% in the two days prior to the first
eruption \citep{Li08}. During which time at least 4 heating (and
cooling) events happened which heated the filament to 1 MK or higher
\citep{Bone}. According to \citet{Bone}, the formation and merger of
the filament which eventually erupted represents 6\% of the total
unsigned flux change during this two day period.

Our model of the heating during the CME eruption and evolution
through interplanetary space will need to match both the charge states
seen in the filament during eruption and detected {\it in situ}, as well as the density
and speed of the ejected material.
\citet{Liewer} present the STEREO EUV images showing the heating of
the filament in the hours and minutes immediately preceding the
eruption. Hot spots are seen in the EUVI 171$\AA$ filter (mainly
Fe IX and XII) at heights near 1.07 R$_{\sun}$.
The magnetic cloud is determined
to have mainly interacted with STEREO B, which penetrated the center
during most of May 22 (DoY 142), while STEREO A only passed through
the magnetic cloud periphery \citep{Kilpua}.
The {\it in situ} Fe charge states seen with PLASTIC on STEREO B and A are shown
in Figure 1 \footnote{Links to the level 2 data are available at http://fiji.sr.unh.edu/ .}.
Charge state data from ACE for multiple ions is shown in Figure 2.
The highest Fe charge states, Fe$^{14+}$, Fe$^{15+}$, are only seen in ACE and STEREO A
which were not centrally located in the event, but rather on its flanks.
Furthermore the appearance of the highest charge states is after the cloud passage as
defined in \citet{Liu08}.

The shock front around the 2007 May 19 CME launched at
958~km~s$^{-1}$, the average speed in transit to 1 AU was around
700~km~s$^{-1}$ and the speed of the MC as it passed the STEREO B
spacecraft was 482~km~s$^{-1}$ \citep{Kilpua}. The filament that
erupted from the active region at about the same time only rose at an
average speed of 103~km~s$^{-1}$ \citep{Liewer}, but it could have
been dragged out by the CME and the fast solar wind to reach a similar
coasting velocity as the ICME. It had no distinct initiation or acceleration
phase, and clearly does not fit within the phenomenology described above and in \citet{RLL07}, which leads us to consider a spheromak model for the CME
evolution.

\section{Resistive Spheromak  model of CMEs}
\subsection{Spheromak Field Configurations}
Spheromaks are  well known force-free configurations of plasma satisfying condition $\nabla\times{\bf B} =\alpha {\bf B}$ with spatially constant $\alpha$.  They are solutions of  the Grad-Shafranov equation in spherical coordinates with  the poloidal current being a linear function of the flux \citep{ChandrasekharKendall57}.
The basic spheromak solution is
\ba &&
B_r = 2 B_0 { j_1\left(\alpha r\right) \over \alpha r} \cos \theta
\nn &&
B_\theta =-B_0 { j_1\left(\alpha r\right) + \alpha r j_1'\left(\alpha r\right) \over \alpha r} \sin \theta
\nn &&
B_\phi = B_0 j_1\left(\alpha r\right) \sin \theta
\label{spheromak}
\ea
where  the spherical Bessel function $j_1$ can be expressed in terms of elementary functions, $j_1(x) = \sin x /x^2 - \cos x /x$. The
parameter $\alpha$ is related to the size of the spheromak $R$, defined by the surface where radial \Bf\ is zero, given by the solution of $j_1=0$; $R = C_\alpha/\alpha$ with $C_\alpha =4.49$.

The spheromak model of CME topology is an attractive
alternative to the usual flux rope model. It is self-contained
with no unknown ``length of flux rope'' parameter to maintain connectivity
to the sun and the heating within it is mathematically
specified once the anomolous resistivity is chosen.
The torus structure, seen on edge, may even resemble the filament shape in some CMEs \citep{Kataoka}.
Previously, \cite{LyutikovKostas10} found {\it ideal} self-similar solutions for {\em  expanding} spheromak, with electric fields
\be
\E =  {r\over c} { \dot{\alpha}  \over \alpha}  {\bf e}_r \times \B
\ee
(dot denotes differentiation with respect to time) and the corresponding {\it non-radial } velocity field
\be
{\bf v}= { \B  (\B \cdot {\bf e}_r)- {\bf e}_r B^2 \over B^2} { r  \partial_t  \ln \alpha }
\label{v}
\ee
In this Section we discuss continuous heating of  an expanding spheromak, generalizing the previous analysis to non-ideal  self-similar expansion with $\E \cdot \B \neq 0$.

Physically, the parallel component of the electric field should be related to the current density through Ohm's law.
Formally, the procedure described below breaks down the assumption of self-similarity, since the value of the parallel \Ef\ component is not linearly  proportional to the current density. Still, we assume that  resistivity plays a subdominant role, so that the expansion remains approximately self-similar. In other words, we assume that resistivity leads to small deviation from the ideal self-similar expansion. The resulting dynamics remain self-similar, by assumption, but are a little different from the ideal case. We are interested  not in the detailed properties of local resistive heating, but in general scaling relations. As we show below, the requirement that an expanding CME dissipates some of its initial energy in the form of heat, detectable by its effect on element charge states by the time it reaches Earth orbit, can be used to estimate the anomalous resistivity.


Expressing \Ef\ in radiation gauge $ {\bf E}= - \partial_t  {\bf A}/c$, where ${\bf A}$
is vector potential,
we find from the force-free condition,
 \be
 \partial_r ( r A_\theta) - \partial_\theta  A_r=
  \alpha r A_\phi  .
 \label{1}
 \ee
This equations highlights two important points. First, it shows how the time-dependences of the poloidal components of the vector potential are related to the toroidal
components, $ A_\theta,A_r \propto \alpha (t) A_\phi$ (this is related to the conservation of poloidal and toroidal fluxes).
Second,
we get one equation for the two functions $  A_r, A_\theta$. Since spheromak solutions are linear, a general solution is a
linear combination of the two solutions.
The two particular solutions, which in a static case correspond to two equivalent choices of the poloidal components of vector potential, in a time varying case  correspond to different physical processes.  Accordingly, there are two types of solutions, corresponding to two choices of vector potential in Eq. (\ref{1}), $A_r \neq 0 $ and $A_\theta \neq 0$. We consider them in turn.

\subsection{Resistive expansion of first  type}

Let us assume that  $B_\phi $ scale with time as an arbitrary function of the expansion parameter, $B_\phi \propto  f(\alpha/\alpha_0)$.  The first type of solutions has $A_r=0$ and, from equation 1,
 \be
 A_\theta =  -B_0      f(\alpha/\alpha_0) { \left(  \sin (\alpha r) - \int ^{\alpha r} {\sin z \over z} dz \right) \over  \alpha^2 r}  \sin \theta .
 \label{A1}
 \ee
The first electromagnetic invariant ${\bf E} \cdot {\bf B} \propto \left(\alpha\dot{f} - 2\dot{\alpha}f\right)$. Thus, if $f= (\alpha/\alpha_0)^2$ the flow is ideal,
consistent with the scaling chosen in  \cite{LyutikovKostas10}.
More generally, if we allow $ f=(\alpha/\alpha_0)^{2+m}$, we find the parallel \Ef\
 \be
 E_{\parallel,1}^2 = m^2  B_0^2 (\alpha/\alpha_0)^{2 m}{\dot{\alpha}^ 2 \sin ^4 \theta\over\alpha _0^4c^2} F_1( r \alpha, \theta)
 \label{E1}
 \ee
where $F_1( r \alpha, \theta)$ is a lengthy function not given explicitly here. The case of $m=0$ corresponds to ideal expansion.
Since we expect that reconnection eliminates magnetic flux, physically realizable solutions correspond to $m>0$. Magnetic flux decreases as
 $\alpha^m \propto R^{-m}$.

 \subsection{Resistive expansion of second type}

The second type of dissipative   expansion corresponds to $A_\theta =0$ and
 \be
 A_r =  B_0 r   f(\alpha/\alpha_0)   j_1 \cos\theta
 \label{A2}
  \ee
  The parallel \Ef\ in this case
   \be
 E_{\parallel,2}^2 =   B_0^2 (\alpha/\alpha_0)^{2(m-1)}{\dot{\alpha}^2\over
 \alpha _0^4c^2}  F_2( r \alpha, \theta).
  \label{E2}
 \ee
For this solution the resulting \Ef\  has a component along \Bf\ for any $m$. The resistive effects in these solutions are induced exclusively  by spheromak expansion. Note, that this   is different from  resistive decay of a stationary spheromak, which proceeds homologously.
Thus, expansion of the spheromak generally leads to appearances of parallel electric fields and associated dissipation, which can proceed much faster than resistive decay of stationary spheromaks.

\subsection{Expansion-driven dissipation}
Let us concentrate on the resistive expansion of first type, which produces equatorial dissipation concentrated in the flux rope, as in the CME observations we attempt to model. The dynamical models described above relate the resistivity to the overall dynamics.
The volume integral over $E_{\Vert}^2$ (see Eq. (\ref{E1})) is
\begin{equation}
\int E_{\Vert}^2dV = 0.033m^2B_0^2R_0^3{v^2\over c^2}\left(R_0\over R\right)^{2m+1}
\end{equation}
which leads to the dissipation rate
\begin{equation}
\dot{\epsilon} =\int {E_{\Vert}^2\over\eta}dV=0.033mB_0^2R_0^3\left(R_0\over R\right)^{2m+1}{v\over R}
\end{equation}
with the identification of the resistivity $\eta = mRv/c^2$. This is justified by integrating the energy dissipation rate above between $R_0$ and $R$ to give a total dissipated energy
\begin{equation}
\int _{R_0}^R{\dot{\epsilon}\over v}dr=0.033{m\over 2\eta}{B_0^2\over c^2}vR_0^4\left(1-\left(R_0\over R\right)^{2m}\right)\simeq 0.033{m\over 2\eta}{B_0^2\over c^2}vR_0^4\times 2m\ln\left(R\over R_0\right)
\end{equation}
where the last step assumes $\left(R-R_0\right)/R_0 << 1$.
With total spheromak (magnetic) energy $U=1.57\times 10^{-2}B_0^2R_0^4/R
\times\left(R_0/R\right)^{2m}$, the difference in energy between $R_0$ and $R$ is $\delta U=U\left(R_0\right)\left(1-\left(R_0/R\right)^{2m+1}\right)$, of which a fraction $2m/\left(2m+1\right)$ is dissipated. Writing
\begin{equation}
\delta U\simeq 2mU\left(R_0\right)\ln\left(R\over R_0\right)=0.033B_0^2R_0^3\ln\left(R\over R_0\right)
\end{equation}
and equating with $\int _{R_0}^R{\dot{\epsilon}\over v}dr$ in equation 11 yields $\eta = mR_0v/c^2$. This is derived assuming an infinitesimal expansion. The generalization to arbitrary expansions gives
\begin{equation}
\eta\simeq mRv/c^2.
\end{equation}

In the second case the dissipation rate
 \be
 \int\E\cdot \J dV = B_0^2 R_0^3 \left({R_0 \over R}\right)^{2 (m +1)} V F_2(m)
 \ee
where $F_2(m) \approx 0.033 m - 5 \times 10^{-4}$, very similar to the first case when integrated over the spheromak volume. The distribution of the heating is different in the two cases, as illustrated in Figure \ref{Diss}.
The amount of energy available for heating
therefore depends on the value of the anomalous resistivity and varies
with position in the spheromak about an average value. The
distribution of volumetric heating rates for the first type of dissipative solutions (of most interest here; see below) is given in Figure \ref{spheromak1}.

To what extent can a spheromak approximate a real CME? Flux rope
structures, similar to the equatorial portion of a spheromak are
frequently observed {\it in situ}, through the rotation of the
magnetic field as the spacecraft penetrates the CME plasma. The
precise origin of the flux rope is not clear. Many authors assume that
such a structure emerges fully formed from the photosphere \citep[e.g.,][]{fan04,magara03}.
Alternatively in the breakout,
\citep[e.g.][]{antiochos99,lynch04}, tether-cutting
\citep[e.g.][]{moore01} or flux cancelation \citep[e.g.][]{linker03}
models the flux rope forms as the CME erupts by the reconnection along
an arcade of magnetic loops.
In such a case, the ends of the flux rope remain attached to the Sun during the eruption. A spheromak however is a self contained magnetic field configuration supported by internal currents, with no external attachment, and presumably if appropriate, must emerge fully formed from beneath the solar photosphere.

In general, the models in which a flux rope forms by reconnection
predict an acceleration phase at the onset of flare reconnection.
 By contrast, the 2007 May 19 CME appears to have moved out from very low down at almost constant speed. There is no obvious feature in its height-time plot that would correspond to the onset of an acceleration phase coupled with reconnection to form a flux rope, suggesting that the observed flux rope must have formed much lower down. Thus its observed trajectory suggests that a spheromak might be a more appropriate description here, than it would have been for other CMEs considered previously in
\citet{RLL07}. \citet{nakagawa10} find a similar trajectory for the magnetic cloud observed by ACE and Nozomi on 1999 April 16-18. They also conclude that the magnetic cloud is better fitted by a toroidal, as opposed to cylindrical, flux rope, though the model fitted is still approximate.
Both flux ropes and spheromaks conserve helicity as they expand, and to do so must decrease their magnetic energy, a part of which can end up dissipated as heat. \citet{kumar96} consider an expanding flux rope, and find temperatures reaching $1 - 2 \times 10^6$K. We apply similar ideas, derived within the context of a spheromak, to the 2007 May 19 CME. We assume that the expansion velocity of the spheromak entering the expression for the heating is given by the observed motion of the CME.

\section{Spheromak non-equilibrium ionization model of the 19 May 2007 CME}

As in \citet{RLL07} we modeled the charge states within the CME ejecta for a variety of
ions using an adaptation of the BLASPHEMER (BLASt Propagation in
Highly EMitting EnviRonment) code
\citep{laming02,laming03b,laming03c}, which follows the time dependent
ionization balance and temperatures of a Lagrangian plasma parcel as
it expands in the solar wind. The fundamental equations are outlined
below.  Further details of the ionization and recombination
calculation can be found in \citet{RLL07} and references therein.

The density $n_{iq}$ of ions of element
$i$ with charge $q$ is given by
\begin{equation} {dn_{iq}\over dt} =
n_e\left(C_{ion,q-1}n_{i~q-1}-C_{ion,q}n_{iq}\right)+
n_e\left(C_{rr,q+1} +C_{dr,q+1}\right)n_{i~q+1}- n_e\left(C_{rr,q}+
C_{dr,q}\right)n_{iq}
\end{equation}
where $C_{ion,q}, C_{rr,q}, C_{dr,q}$ are the rates for electron
impact ionization, radiative recombination and dielectronic
recombination respectively, out of the charge state $q$. These rates
are the same as those used in the recent ionization balance
calculations of \citet{bryans06}, with more recent updates given in
\citet{RLL07}. The electron density $n_e$ is
determined from the condition that the plasma be electrically
neutral. The ion and electron temperatures, $T_{iq}$ and $T_e$ are
coupled by Coulomb collisions by
\begin{equation} {dT_{iq}\over dt}= -0.13n_e{\left(T_{iq}-T_e\right)\over M_{iq}T_e^{3/2}}
{q^3n_{iq}/\left(q+1\right)\over\left(\sum _{iq}
n_{iq}\right)}\left(\ln\Lambda\over 37\right)
\end{equation}
 and
\begin{equation}
{dT_e\over dt}= {0.13n_e\over T_e^{3/2}}\sum
_{iq}{\left(T_{iq}-T_e\right)\over M_{iq}}
{q^2n_{iq}/\left(q+1\right)\over\left(\sum _{iq}
n_{iq}\right)}\left(\ln\Lambda\over 37\right) -{T_e\over
n_e}\left({dn_e\over dt}\right)_{ion} - {2\over 3n_ek_{\rm B}}
{dQ\over dt}+{\dot{\epsilon}\over 1.5n_ek_{\rm B}}
\end{equation}
Here $M_{iq}$ is the atomic mass of the ions of element $i$ and charge
$q$ in the plasma, and $\ln\Lambda\simeq 28$ is the Coulomb
logarithm. The term in $dQ/dT$ represents plasma energy losses due to
ionization and radiation. Radiation losses can be taken from
\citet{summers79}. In a CME from a filament eruption the densities and
temperatures are such that radiative losses can be important, unlike
most applications in the solar wind where they are generally negligible.
The term
$-\left(T_e/n_e\right)\left(dn_e/dt\right)_{ion}$ gives the reduction
in electron temperature when the electron density increases due to
ionization. Recombinations, which reduce the electron density, do not
result in an increase in the electron temperature in low density
plasmas, since the energy of the recombined electron is radiated away
(in either radiative or dielectronic recombination), rather than being
shared with the other plasma electrons as would be the case for
three-body recombination in dense plasmas.
We include the last electron heating term to model the Ohmic dissipation given by equation 10 for the heating in the spheromak.

In addition to ionization and recombination there is
also the geometry to consider in the density and temperature
evolution. For lack of a better motivated profile we chose a
uniform initial density and temperature, exploding from an initial
radius of $(z_0-1)$ where $z_o$ is the starting height from the center of
the sun in solar radii. Thus the underlying density evolution for
adiabatic expansion goes as $[(z-1)/(z_0 -1)]^2$, which goes over to
$\left(z/z_0\right)^2$ as $z >> z_0$ and the CME expansion more closely follows that of the ambient solar wind.

There are two questions we are trying to answer with our
simulations. First, how well can a spheromak solution
explain the charge states seen in the 19 May 2007 event?
To this end, two sets of models were explored. In model 1 we attempted to match both the filament
heating and the ICME charge states using only the heating due to anomalous resistivity
in the spheromak. In model 2 we allowed the plasma to already have been heated
during the eruption to temperatures that would explain the EUVI observations, but
then followed the subsequent heating in the spheromak. Second, given that fast CMEs, particularly at solar
maximum, often show Fe charge states as high as Fe$^{17+}$ {\it in situ}, what
are the highest Fe charge states obtainable by anomalous resistive
heating in a spheromak model? In other words, how general could such a model be?

The baseline model 1 starts the plasma at a temperature of $10^{4}$K,
a density of $10^{10}$cm$^{-3}$, and a velocity of 10~km~s$^{-1}$
accelerating to 500~km~s$^{-1}$. These were chosen as appropriate
temperatures and densities for cool filament material, and the
evolution is consistent with the final density seen {\it in situ}.  We
initially explored the effect on the final charge states of changing
the anomalous resistivity, and thus the amount of magnetic flux
dissipation, through the parameter $m$.  We assumed heating at the
average rate for the spheromak volume.
Values of $m$ of 0.01 and below leave the charge states
essentially unaffected by the spheromak heating. (The charge states
still evolve with the adiabatic expansion which changes the
temperature and density as the CME erupts). At $m$ values between 1.5
and 2 one reaches the limit of diminishing returns whereby
increasing $m$ does not result in higher charge states, due to the factor
$\left(R_0/R\right)^{2m}$ in equation 10. This corresponds to an anomalous
resistivity between $8\times 10^{-5}$s and $10^{-4}$s, as much as
$10^{11}$ times the Spitzer value, with a maximum Fe charge state of Fe$^{13+}$.
We note that increasing the final velocity or the acceleration rate
does not substantially change this limit. A more dramatic spheromak
expansion starting from a smaller sphere does heat the material
faster, but it also reduces the density too quickly such that (a) the
final densities are substantially lower than seen {\it in situ} and (b)
ionization declines such that the highest charge states are not
produced. Thus we conclude that the high charge states over Fe$^{13+}$ in
fast CMEs cannot be reproduced purely with the
spheromak solution and anomalous resistivity. This leaves the question of how
well the spheromak can explain this particular 19 May 2007 CME.

We choose $m=0.75$ and compute ion charge states averaged over the whole spheromak assuming the distribution of heating rates in Figure 3.
Since the 2007 May 22 CME flux rope does not exhibit Fe more ionized than Fe$^{13+}$,
its highest charge states can be explained with the baseline model 1
and an average $m=0.75$. The overall weighted charge states
of this model are shown in Figure 5. There are two obvious
discrepancies between {\it in situ} observations and the Fe charge states
modeled here. First, there are too
many low charge states from the unheated portion of the spheromak.
This is not a major concern as it is possible that that portion of the
spheromak is simply not filled with plasma. Second, Carbon charge states seem to indicate that the plasma started hotter than in these simulations.

Additionally one can consider whether the spheromak heating could
explain the EUV observations of the filament around launch.
Figure 6 shows the early evolution of the
charge states corresponding to the average heating rate near the sun in the baseline model 1 with $m=0.75$. This started with the
plasma at 1.1$R_{\sun}$, above the heights of 1.07$R_{\sun}$ where
\citet{Liewer} first see Fe IX and Fe XII emission. Starting simulations below this height, with a density that evolves to become the observed density {\it in situ} at earth orbit, leads to strong radiative cooling that keeps the electron temperature below $\sim 10^4$K until a radius of 1.1$R_{\sun}$ is reached, whereupon the density is sufficiently reduced to lengthen the radiative cooling time. Hence the heating observed by \citet{Liewer}, beginning before the actual eruption began, must have had a different origin.

Testing models that started at lower
heights, we found that indeed Fe IX and even Fe XII could be
produced within a short distance of launch. We did not use these for
our {\it in situ} modeling because for reasonable starting filament
densities the final density due to the rapid expansion was too low.
Overall we conclude that while reproducing the EUV observations may be
possible within a spheromak context, given the sequence of heating and
cooling of the filament leading up the event as described by
\citet{Bone}, it is also plausible, and probably more likely, that whatever mechanism was heating the filament pre-eruption was responsible for the filament heating during the eruption.

The above discussion naturally leads us to our second model to
determine if spheromak heating starting from temperatures seen in EUV
observations can explain the {\it in situ} charge states. In Model 2 we
started the plasma at $10^{6}$K, $10^{9}$cm$^{-3}$, and a velocity of
100~km~s$^{-1}$ similar to the velocity of the filament from
\citep{Liewer} allowing it to then accelerate to the coasting velocity
of 500~km~s$^{-1}$. Starting from this high temperature, the
additional heating of the spheromak has no noticeable effect on the
charge states. While the temperature was chosen such that the charge
states at low heights match observations, the final Fe charge states
fall short of those detected {\it in situ}. We also experimented with
starting at $10^{6}$K but the higher density and/or at a slower
velocity. However, radiative recombination dominated bringing the
charge states and temperature rapidly back to their $10^{4}$K levels.
This has implications for the rapidity and amount of energy that had
to be dumped into the filaments to produce their high charge states. A
thorough examination of this phenomenology is warranted but we leave
this to future work.
Overall, starting from a higher initial temperature in a uniformly
filled spheromak did not solve the
problem of matching the highest Fe charge states.

\section{Discussion and Conclusions}
We have explored modeling this filament eruption as due to the heating
provided in an expanding spheromak. Figure 6 shows the evolution of H, C, O,
Si and Fe charge states obtained with an anomalous resistivity
increased from the classical (i.e. Spitzer) value by a factor around
$2\times 10^7 - 6\times 10^{11}$, as the plasma temperature varies between
$10^4$ K and $10^7$ K. Anomalous resistivity is thought to arise in conditions where plasma turbulence can scatter current carrying electrons, and can do so more effectively than the Coulomb collisions between electrons and ions. It may be estimated by replacing the isotropization frequency by Coulomb collisions with the relevant wave-particle interaction rate.

Solutions of the one dimensional Vlasov equation designed to model the Buneman instability in a reconnecting current sheet \citep{Wu09}, and therefore model electrons in Langmuir turbulence,
give quiescent values of the anomalous resistivity of $\times 10^7$ the Spitzer value, with transient values as
high as $\times 10^9$ in a background plasma temperature of $10^7$
K. In our case, dropping the anomalous resistivity to $\times 10^{10}$ the classical
value, the dominant Fe charge state becomes Fe$^{7+}$. The ratio of
the electron plasma frequency to the collision frequency at $1.4
R_{\sun}$ in Figure 6 evaluates to $\sim 10^9$, and probably
represents the maximum enhancement in resistivity than one may
reasonably expect. \citet{lin07} estimate a much higher resistivity from the observed width of current sheets trailing CMEs, as high as $10^{12}$ times the classical resistivity. \citet{bemporad08} revisits this and argues that what is observed as a macroscopic current sheet is in fact an assembly of microscopic current sheets, each with width corresponding to anomalous resistivities similar to those given by \citep{Wu09}, although possibly attributed to different modes of turbulence.  We therefore
conclude that magnetic energy dissipated in the expansion of a
spheromak may account for the filament heating between $10^4$ K and
$10^6$ K, but is unlikely to be the explanation for the high Fe charge
states characteristic of temperatures of order $10^7$ K and
higher. The nature of the heating specified in the spheromak solution
is such that the heating rate due to anomalous resistivity cannot be
increased without limit, even if an anomalous resistivity as high as
allowed by the solution could be justified physically.

Higher charge states, beyond Fe$^{13+}$ must therefore result from other means of heating. Noting that in the 2007 May 19 CME, the highest Fe charge states are observed exterior to the flux ropes, we speculate that reconnection associated with the eruption is most likely. Type II and Type III radio bursts were observed during this eruption \citep{Kerdraon} at times when the CME shock was between 1.2 and 1.4 $R_{\sun}$ heliocentric distance \citep{Veronig}. Bearing in mind that the plasma we observe and model will be at lower altitude that the CME shock, this corresponds very well to the epochs of heating shown in Figure 6. Recent results suggest that reconnection might be an efficient means of heating electrons \citep{drake06,oka10}.

In earlier work \citep{RLL07}, we expressed the heat input to the CME in terms of its kinetic energy input by relating the heating to the acceleration, tacitly assuming that the mechanism of acceleration inevitably also supplies heat to the CME plasma. In the case of the spheromak, the heating is related to the expansion, but not the acceleration, so the ratio of heat to kinetic energy does not have a natural interpretation. We estimate it, though, to facilitate comparison with \citet{RLL07}. From equations 11 and 12,
\begin{equation}
{\int\dot{\epsilon}dt\over mv^2/2}={0.033B_0^2R^3\ln\left(R/R_0\right)\over
2\pi\rho R^3v^2/3}= 0.2{v_A^2\over v^2}\ln\left(R\over R_0\right),
\end{equation}
which close to the Sun evaluates to $\sim 1$. This is similar to, but slightly lower than typical values found by \citet{RLL07}. The CMEs studied in that work frequently had charge states of Fe up to neon-like, rather higher than in the flux rope studied here which appear to be consistent with our estimate of thermal to kinetic energy in the CME.

\acknowledgements This work has been supported by NASA Contract NNG08EK62I and by basic research funds of the Office of Naval Research.

\begin{figure}[h]
\centerline{\includegraphics[width=3.25in]{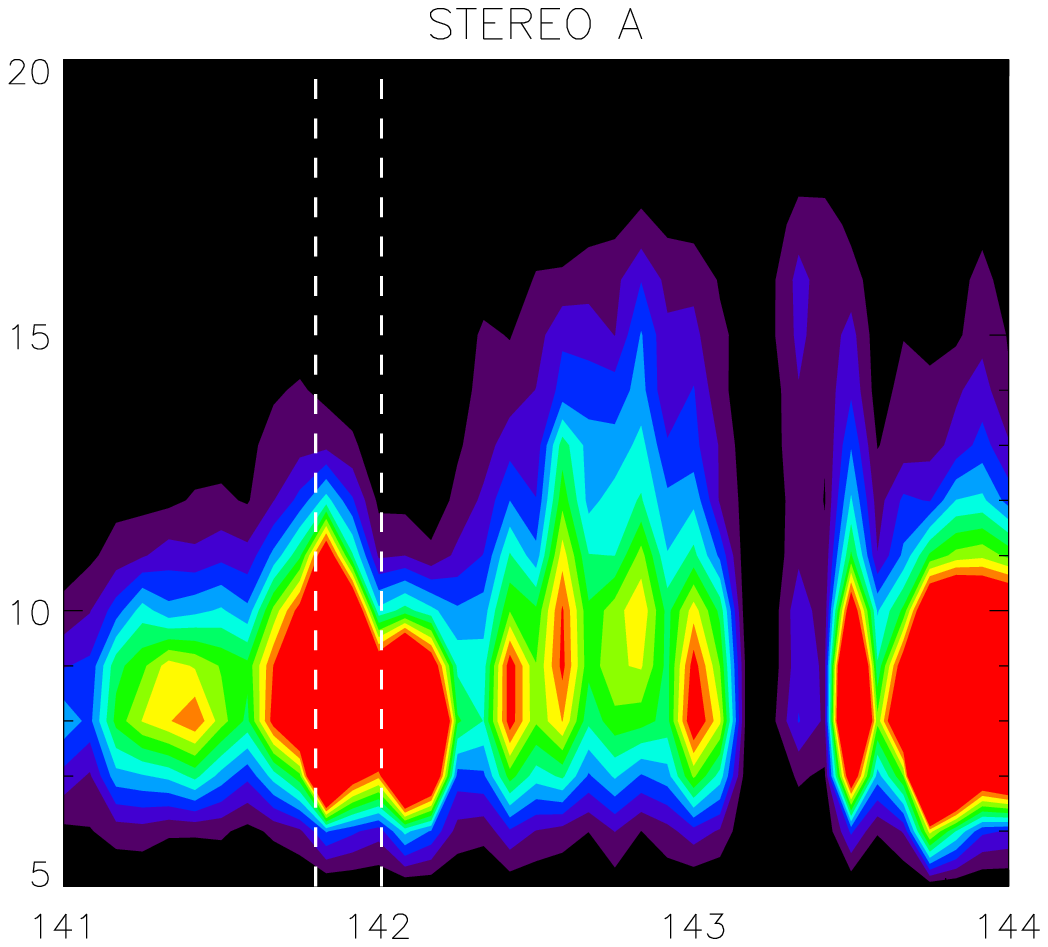}
\includegraphics[width=3.25in]{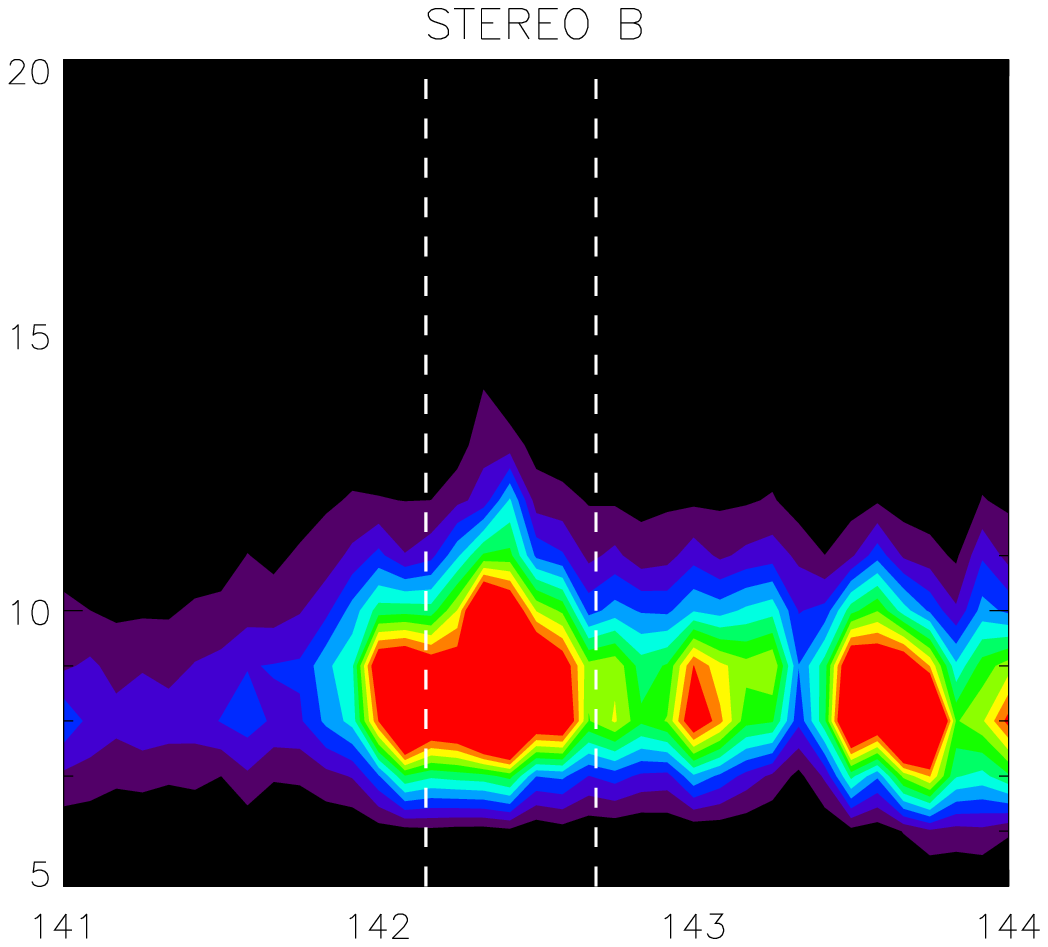}}
\caption{Fe charge state distributions detected by STEREO A and B during
the 2007 May 19 CME passage. Highest Fe charge states are detected by STEREO A,
though the magnetic cloud is more clearly detected in STEREO B. Magnetic field
and other data indicate cloud passage between 3:36 hr and 16:34 hr on day 142 (22 May)
for STEREO B and 21 May 19:12 until 22 May 00:14 for STEREO A \citep{Liu08}. The dashed vertical white lines show the temporal extent of the magnetic cloud (or flux rope) in this event.}
\end{figure}

\begin{figure}[h]
\includegraphics[width=3.25in]{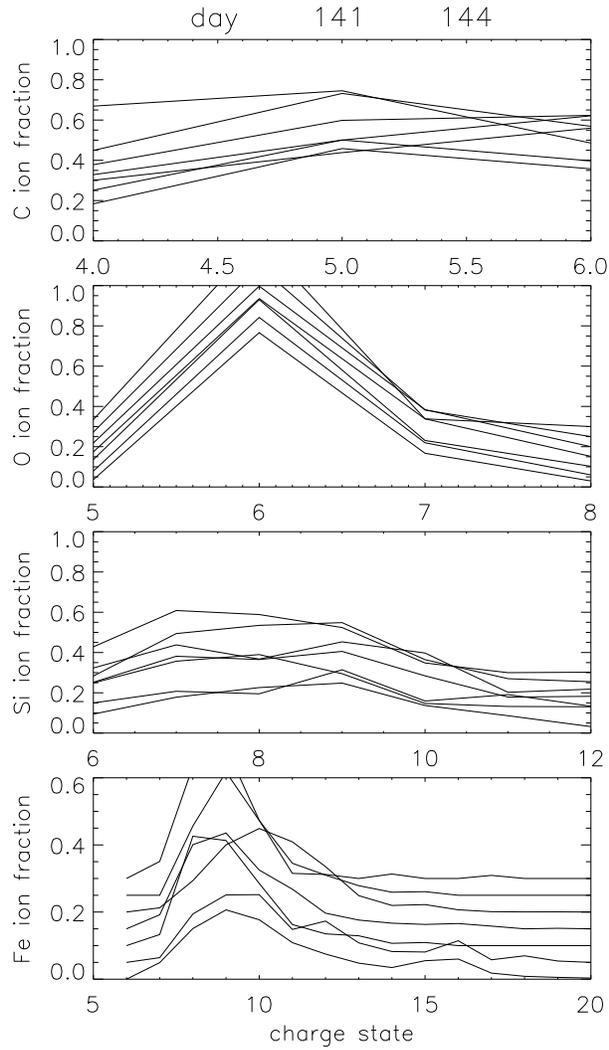}
\caption{ACE charge state data for C, O, Si and Fe in 8 hr bins. Earliest times
have been offset upwards for display purposes. The highest Fe charge states only
appear starting in Day 143 after the passage of the interior of the magnetic cloud.
\citet{Liu08} report the ICME passages through ACE as 21 May 22:19 to 22 May 12:43.}
\end{figure}

\begin{figure}[h]
   \includegraphics[width=0.55\linewidth]{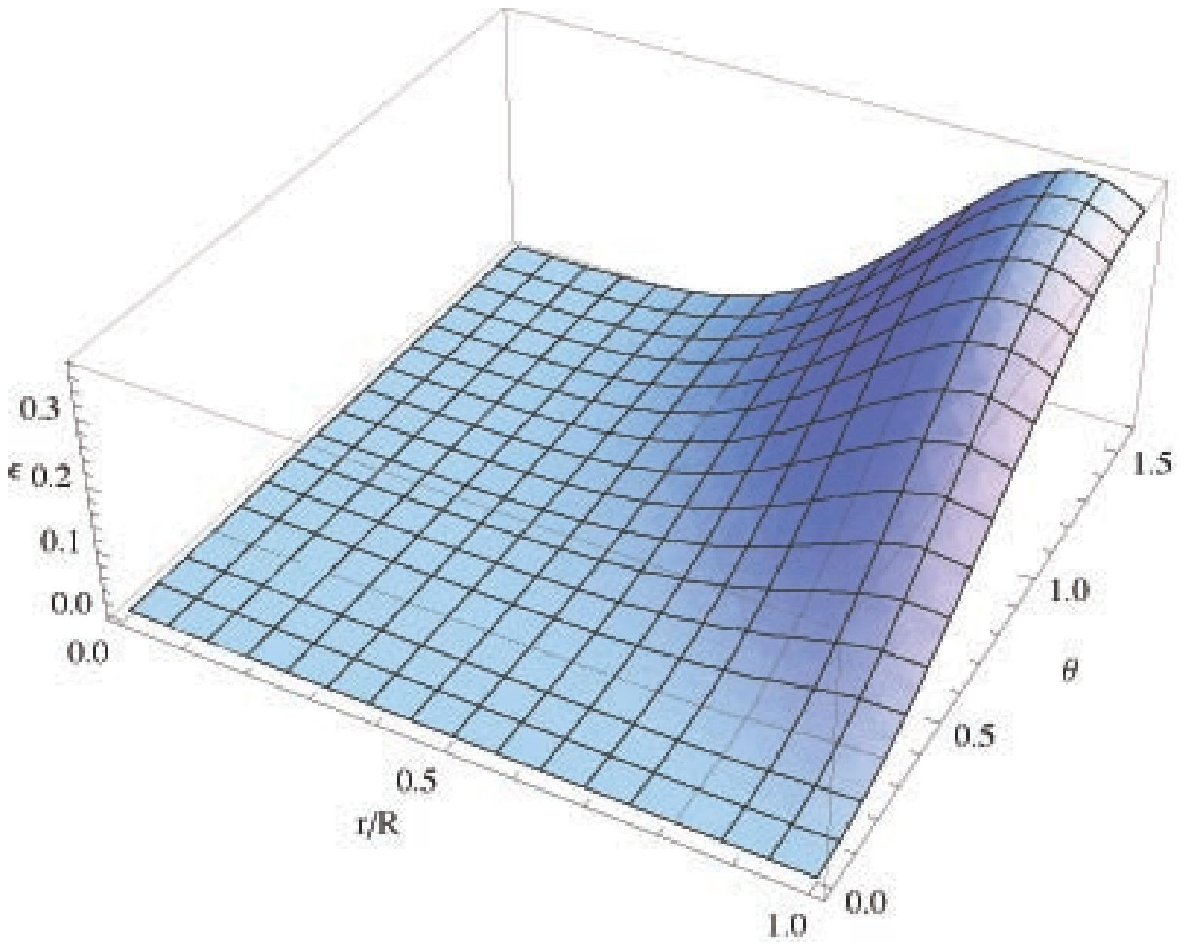}
     \includegraphics[width=0.55\linewidth]{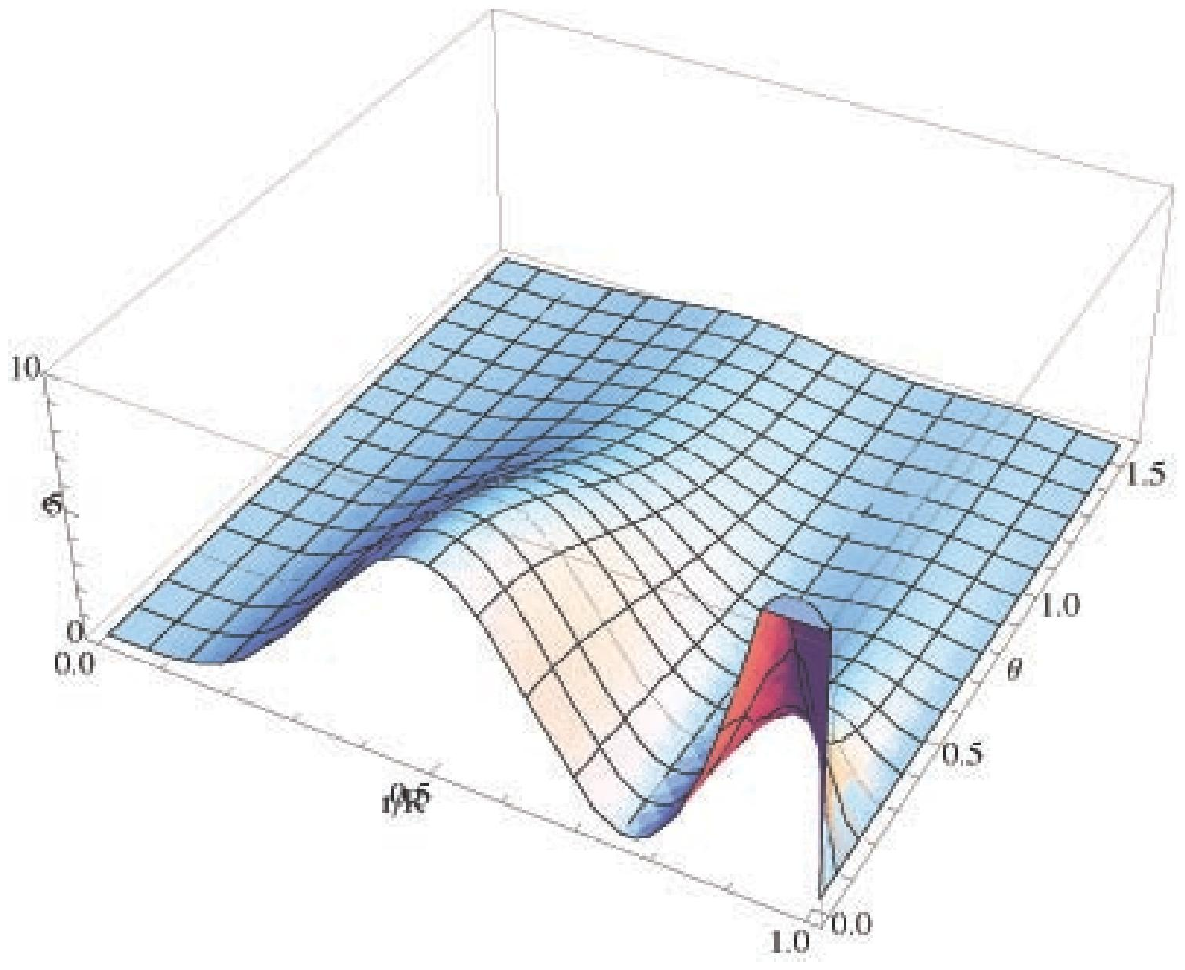}
   \caption {Normalized dissipation rate $\epsilon$ as function of radial distance $r/R(t)$ and polar angle $\theta$ for two self-similar dissipative structures, m=1. {\it Left Panel}:  solutions  (\ref{A1}-\ref{E1}). {\it Right Panel}:  solutions (\ref{A2}-\ref{E2}).
   }
 \label{Diss}
 \end{figure}

\begin{figure}[t]
\includegraphics[width=6in]{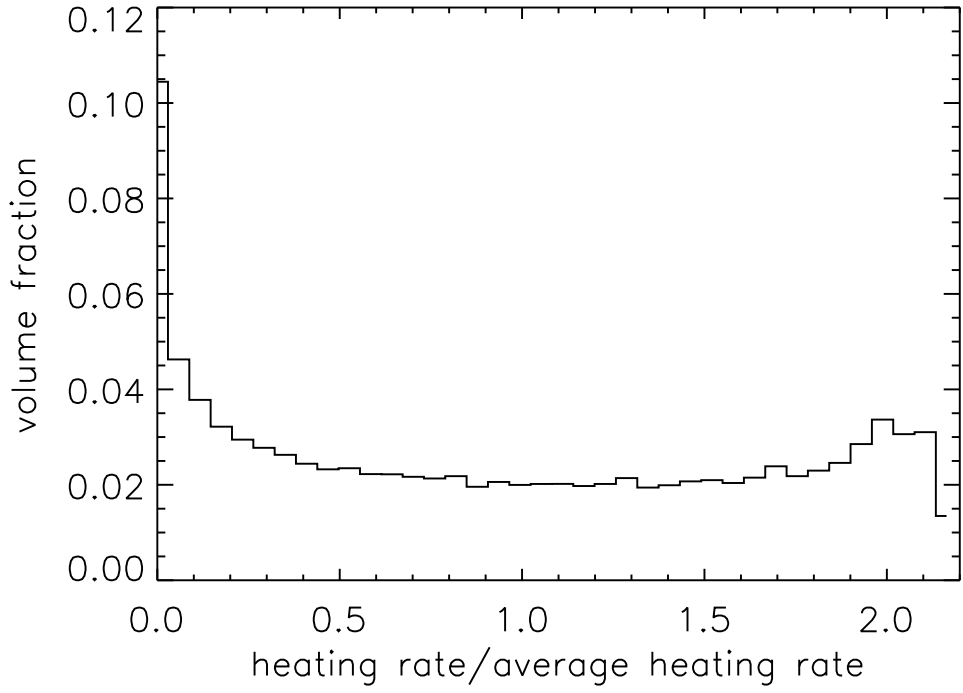}
\caption{Distribution of volumetric heating rates in the spheromak
  model of the first type, solution (\ref{A1}-\ref{E1}),
in terms of the average heating rate over the whole volume.}
\label{spheromak1}
\end{figure}

\begin{figure}[h]
\includegraphics[width=6.5in]{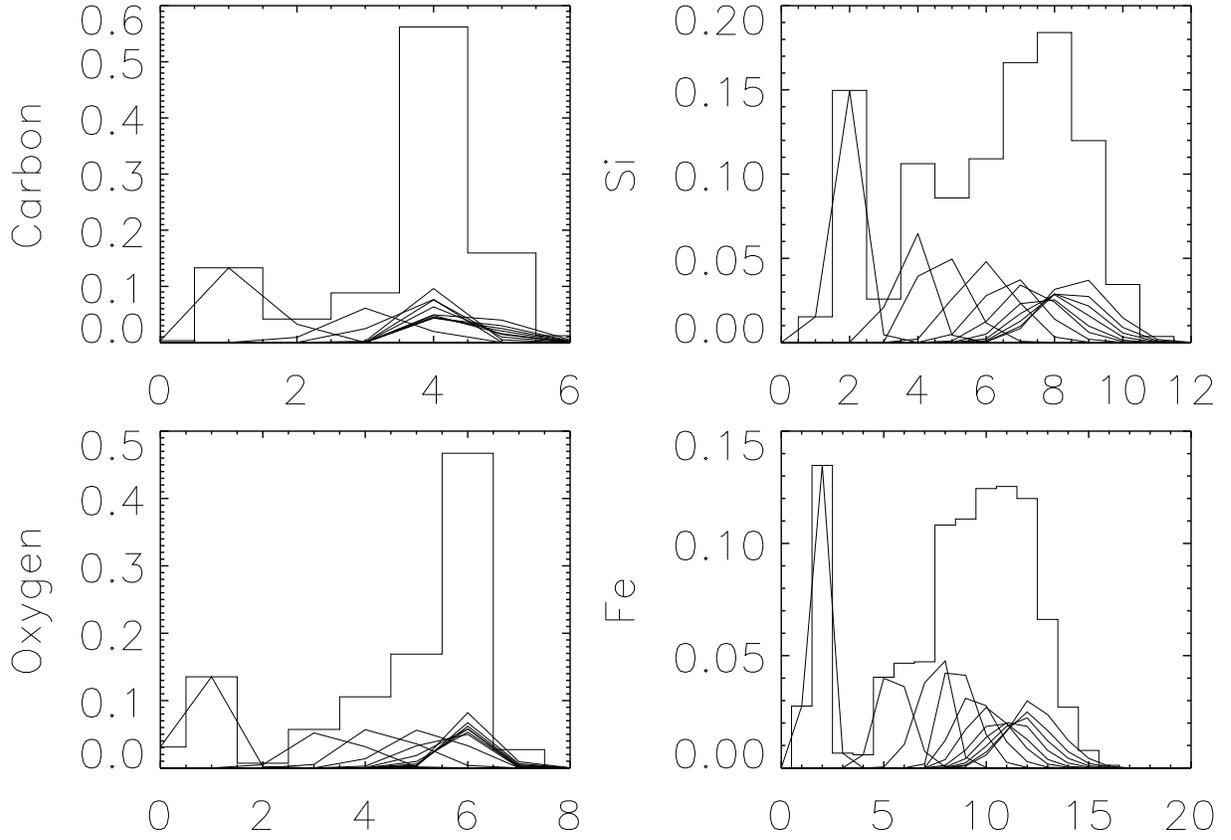}
\caption{Final charge state distributions of C, O, Si and Fe for model 1.
The individual lines show the contributions from regions within the spheromak with different heating rates.
The histograms show the cumulative charge distribution assuming a volume weighted summation. }
\end{figure}

\begin{figure}[h]
\includegraphics[width=6.5in]{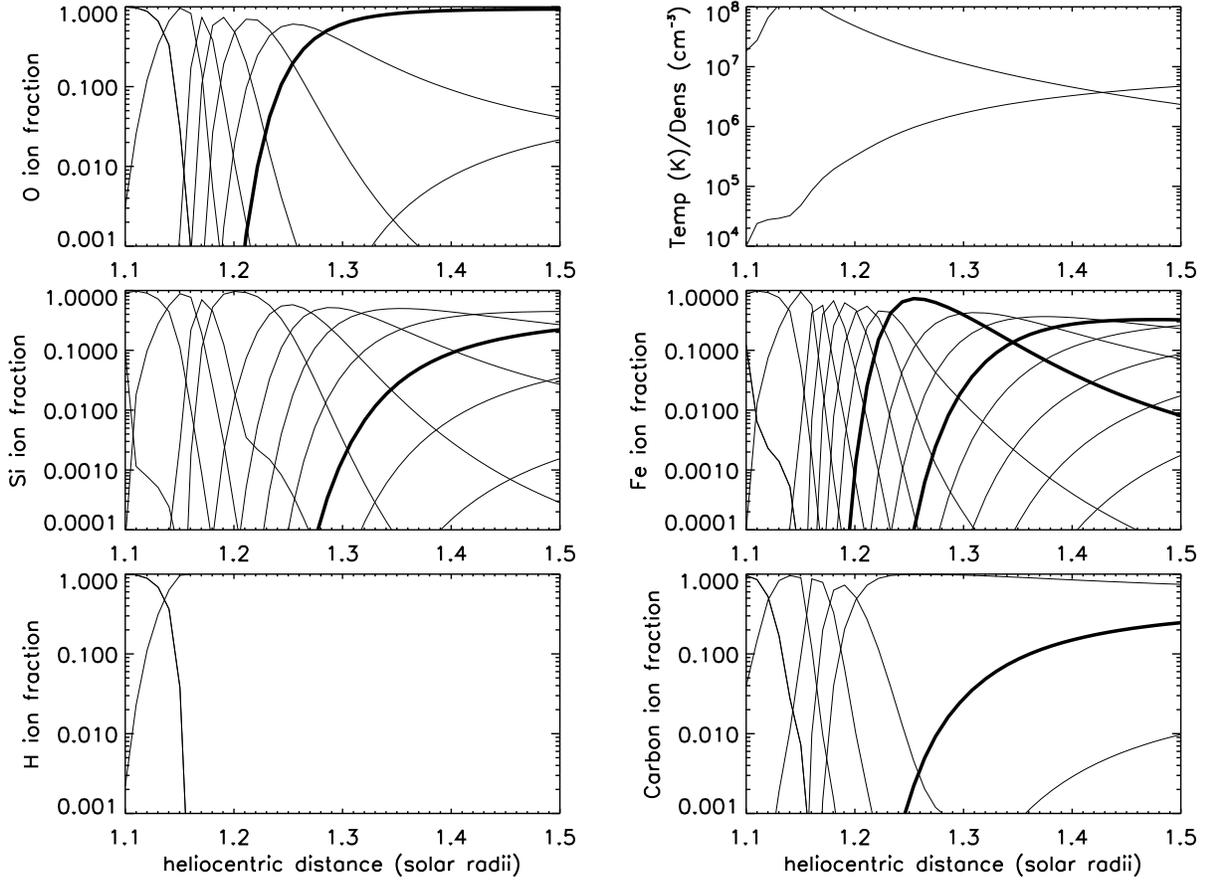}
\caption{Evolution of charge state distributions for the highest anomalous
resistivity considered in model 1 for H, C, O, Si, and Fe. Thick lines
highlight O$^{+6}$, Si$^{+9}$, Fe$^{+8}$, Fe$^{+11}$, C$^{+5}$.}
\end{figure}
\end{document}